# Polytropic behavior in the structures of Interplanetary Coronal Mass Ejections

M. A. Dayeh[1,2] and G. Livadiotis[3]

[1] *Southwest Research Institute, San Antonio, TX  78238 (maldayeh@swri.org)*

[2] *University of Texas at San Antonio, San Antonio, TX 78249*

[3] *Princeton University, Princeton, NJ, 08544*

**Abstract.** The polytropic process characterizes the thermodynamics of space plasma particle populations. The polytropic index, $\gamma$, is particularly important as it describes the thermodynamic behavior of the system by quantifying the changes in temperature as the system is compressed or expanded. Using Wind spacecraft plasma and magnetic field data during 01/1995 – 12/2018, we investigate the thermodynamic evolution in 336 Interplanetary Coronal Mass Ejection (ICME) events. For each event, we derive the index $\gamma$ in the sheath and magnetic ejecta structures, along with the pre- and post- event regions. We then examine the distributions of all $\gamma$ indices in these four regions and derive the entropic gradient of each, which is indicative of the ambient heating. We find that in the ICME sheath region, where wave turbulence is expected to be highest, the thermodynamics takes longest to recover into the original quasi-adiabatic process, while it recovers faster in the quieter ejecta region. This pattern creates a thermodynamic cycle, featuring a near adiabatic value $\gamma \sim \gamma_a$ (=5/3) upstream of the ICMEs, $\gamma_a - \gamma \sim 0.26$ in the sheaths, $\gamma_a - \gamma \sim 0.13$ in the ICME ejecta, and recovers again to $\gamma \sim \gamma_a$ after the passage of the ICME. These results expose the turbulent heating rates in the ICME plasma: the lower the polytropic index from its adiabatic value and closer to its isothermal value, the larger the entropic gradient, and thus, the rate of turbulent heating that heats the ICME plasma.

## 1. Introduction

Coronal Mass Ejections (CMEs) are gigantic eruptions of magnetized plasma structures from the Sun (e.g., Gopalswamy 2006; Webb & Howard 2012, Dayeh 2017, Kilpua et al. 2019) that propagate in the interplanetary (IP) space to reach far into the outer heliosphere (Wang et al. 2001, Richardson et al. 2003). Once away from the Sun, CMEs evolve into interplanetary CMEs (ICMEs). ICMEs that are sufficiently faster than the ambient solar wind tend to drive an IP shock.

These IP shocks can accelerate IP particles to high energies, often contributing to solar energetic particle (SEP) events associated with CME eruptions. A nominal ICME comprises two major parts: (i) Sheath – a region of compressed, turbulent, and inhomogeneous plasma that exhibits high variability of plasma physical processes (e.g., Kilpua et al. 2019) and (ii) Magnetic Obstacle (MO) - a solar wind structure that is often associated with enhanced magnetic field, smooth rotation, and a suppressed proton temperature (e.g., Burlaga et al. 1981). ICME sheaths are key drivers of geomagnetic activity at Earth, leading to intense space weather consequences (Tsurutani et al. 1988). Enhanced sheath turbulence can particularly intensify the geomagnetic activity (e.g., Borovsky & Funsten 2003). Understanding ICME propagation and internal processes is thus an important aspect to further advance our space weather capabilities.

Remote observations of CMEs close to the Sun have provided insights into the heating and acceleration processes of the embedded plasma. Studies using SOHO spacecraft extreme ultraviolet (EUV) measurements showed that CMEs ambient temperatures are higher than the ambient solar wind, suggesting that the inner corona transfers thermal energy into the CME plasma, where CMEs have higher temperatures than the ambient solar wind (Kohl et al. 2006; Bemporad & Mancuso 2010; Mishra et al. 2020). On the other hand, in situ measurements of ICMEs have provided a plethora of information about ICME plasma properties and evolution. Compared to the ambient solar wind, in situ observations from missions across the heliosphere (Voyagers:>1 au - >100 au; Ulysses:5 au; Helios:<1 au; WIND:1 au; ACE:1 au; and STEREO:1au) provided direct evidence that ICMEs have lower temperatures (Burlaga et al. 1981; Richardson & Cane 1993) and highly elevated ionic charge states (Lepri et al., 2001; Zurbuchen et al., 2003). The latter is an indicative of strong heating at the CME source in the lower corona.

Thermodynamic properties of ICMEs can be studied using a polytropic state estimation. Polytropic behavior defines specific relations among thermodynamical observables such as plasma density and temperature. In numerous cases, space plasma density and temperatures positively correlate with different values of the polytropic index, γ, and thus indicating different thermodynamical states (e.g., Totten et al. 1995; Nicolaou et al. 2014; Livadiotis & Desai 2016).

The polytropic relationship may be expressed by

$$P \cdot n^{-\gamma} = T \cdot n^{1-\gamma} = const.,\qquad(1)$$

that involves the thermal pressure *P*, the temperature *T*, and the density *n* of the plasma. The polytropic expression for the logarithms of the thermal variables becomes linear and allows a straightforward estimation of the polytropic index *γ* by examining datasets of log(*T*) vs. log(*n*).

Variations in γ among ICME structures is an indicative of heating processes. Evolution of these processes in time and space as the ICME expands (e.g., Lugaz et al. 2020) can be studied in radially-aligned spacecraft over different distances in the heliosphere (e.g., Phillips et al. 1995; Skoug et al. 2000). It is important to note that thermodynamical properties as functions of radial distance from the Sun is a key approach to understand the ICME physical properties and heating processes. Thus, analysis of an ensemble is required to draw physical conclusions on the evolution of the thermodynamical properties from a single location.

Statistical studies at different distances from the sun have showed that as a function of radial distance away from the Sun and compared to the ambient solar wind, proton density and magnetic field magnitudes decrease at a faster rate in ICMEs, while the temperature drops at a slower rate, indicating that the expansion of ICMEs in the IP space is closer to an isothermal process (e.g., Wang & Richardson 2004; Liu et al. 2005; 2006; Wang et al. 2005; Mishra et al. 2020).

In this work, we examine the polytropic behavior and the change of entropy in four different regions, representing before, during, and after the passage of 336 ICMEs measured at 1 au, and then, connect the evolution of the polytropic behavior and the corresponding thermodynamic states to the heating of the ICME plasma. We find that ICMEs exhibit a thermodynamic cyclic behavior that expose the turbulent heating rates in the ICME; the lower the polytropic index from its adiabatic value and closer to its isothermal one, the larger the rate of turbulent energy that heats the ICME plasma. We organize the letter as follows. Section 2 and 3 describe the data used and the methodology followed by our analysis. The results are shown in Section 4. Finally, Section 5 summarizes and discusses the results.

## 2. Data

We use the following datasets: (i) Wind mission ICME catalogue (Nieves-Chinchilla et al. 2018) between 1995 to 2018, with 336 ICMEs in total; (ii) high cadence solar wind plasma and magnetic-field measurements from the Solar Wind Experiment (SWE; Ogilvie et al. 1995) and the Magnetic Field Investigation (MFI; Lepping et al. 1995) onboard Wind; and, (iii) the derived plasma moments from Wind measurement provided by OMNI Coordinated Data Analysis Web (CDAWeb).

Figure 1 shows the four regions of interest, that is, in the ICME sheath and magnetic ejecta structures, along with the pre- and post- event regions. The sharp jump in the magnetic field indicates the arrival of the ICME-driven shock followed by the magnetically turbulent sheath region, which in turn is followed by the magnetic obstacle demonstrated by the slow rotation of the magnetic field.

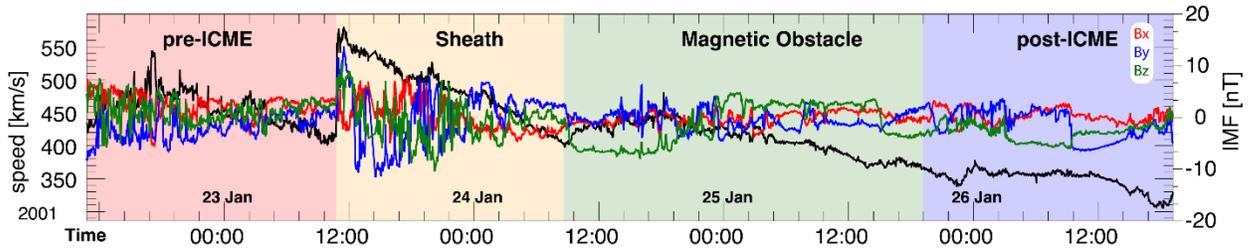

**Figure 1.** Solar Wind (SW) speed (black) and interplanetary magnetic field (IMF) vector ($B_x$: red, $B_y$: blue, $B_z$: green) time series showing an example of an ICME and the four examined regions, pre-ICME SW (light red), sheath (light brown), magnetic obstacle / ejecta (light green), and post-ICME SW (purple).

## 3. Methodology

The methodology of deriving the polytropic indices is summarized in the following steps, which are also illustrated in Figure 2. Details of this method and its sensitivity has also been substantiated in previous analyses, e.g., Kartalev et al. 2006; Nicolaou et al. 2014; Nicolaou & Livadiotis 2019.

1. *Time intervals:* We use a sequence of 5-minute moving window comprising five data points. This approach is validated and aims at minimizing mixed particle measurements on different streamlines (Figure 2a).
2. *Data filtration:* For each selected interval, we examine the stability of Bernoulli's integral by requiring the variance over the mean to be less than 10%. This condition is particularly important, as it ensures that γ is derived for-likely-the same plasma ensemble along the same streamline. (Figure 2a, not shown).

3. *Density-Temperature fitting:* We fit a linear model to the data plot of log(*T*) vs. log(*n*) to infer the value of *γ* (Equation 1; Figure 2b).
4. *Determining γ*: Repeating step 3 for all time windows in any region results in a set of polytropic indices, from which the mean can then be determined (Figure 2c).

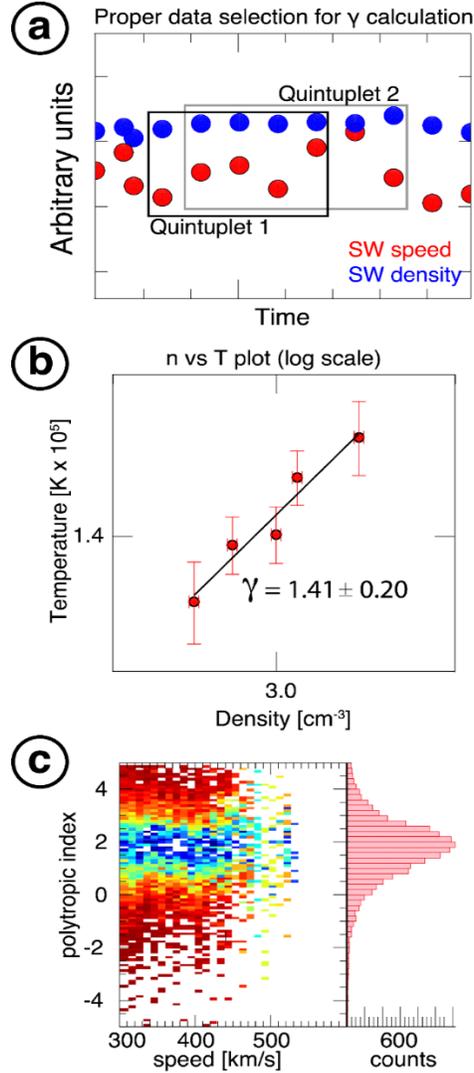

**Figure 2.** (a) Illustration of the solar wind selection when calculating the polytropic index (see text for details). (b) Temperature vs. density plotted on a log-log scale, where the polytropic index is given as slope = $\gamma-1$. (c) Derived polytropic indices are plotted against the solar wind speed in a 2-dimensional histogram (normalized according to Livadiotis & Desai 2016) (right).

We derive the polytropic index in four regions associated with each ICME event, namely, pre-ICME (1 day prior to shock arrival), Sheath, Magnetic obstacle, and post-ICME (1 day after), as illustrated in Figure 1. We further constrain the selection of the events to comply with the following criteria: (i) All the examined four regions (pre-ICME solar wind, sheath, magnetic obstacle / ejecta, and post-ICME solar wind) have to be characterized by a derived value of the polytropic index $\gamma$; then, the polytropic indices of a region are only accepted if they are statistically significant: more than 10 derived values of polytropic index, with standard deviation less than the 20% of the respective mean; there must be no statistical significant trend of the derived polytropic indices with speed (e.g., Livadiotis & Desai 2016; Livadiotis 2018; Livadiotis et al. 2018). Among all the studied events (336), we have selected 25 'clean' events that follow the aforementioned criteria.

## 4. Results

Figure 3 shows the histograms of all the derived polytropic indices in the four studied regions (green), and those of the 'clean' selected events (yellow). The enriched histograms of the selected events are also shown. The enriched histograms are specifically constructed so that both the binned values and their uncertainties are taken into account. (Details on the technique are described in Livadiotis 2016).

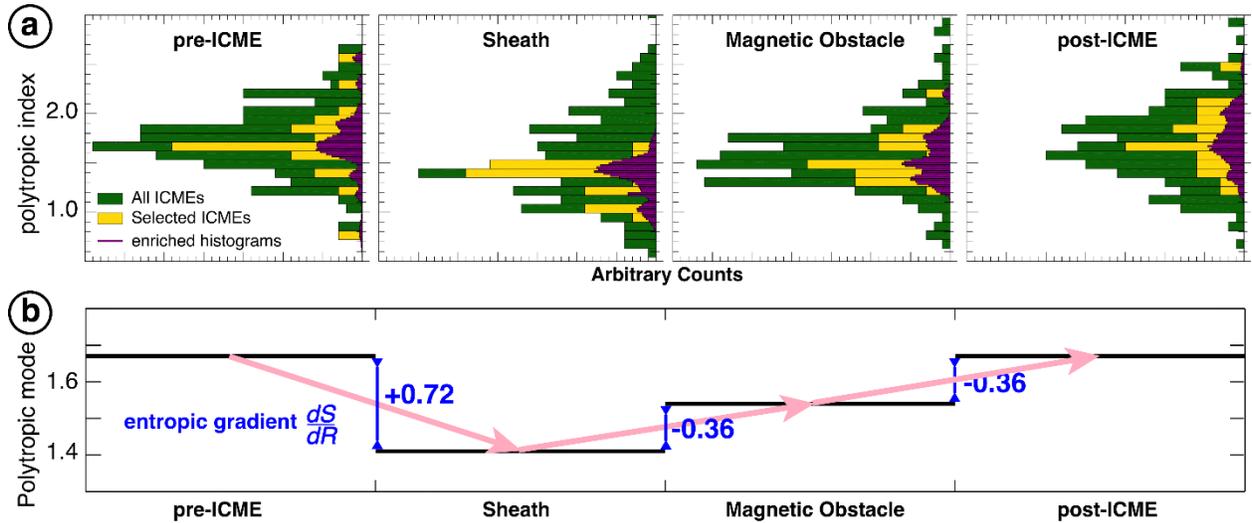

**Figure 3.** (a) Histograms of all the events (green), of the selected events (yellow), and of the enriched dataset of the good events (purple), all plotted for each region. (b) Calculated entropic gradient for $R=1$ AU according to Eq.(6).

Next, we connect the derived polytropic indices with the entropy gradient (Adhikari et al. 2020; Livadiotis et al. 2020) and the turbulent heating gradient (Verma et al. 1995; Vasquez et al. 2007; Livadiotis 2019; 2021). Combining the entropy expressed in terms of temperature and density,

$$\tfrac{1}{k_{\rm B}} S = \ln W \text{ with } W \propto T^{d/2} / n, \qquad (2)$$

the polytropic relationship between temperature and density

$$T \propto n^{\gamma-1}, \qquad (3)$$

and the density drop due to the CME expansion

$$n \propto R^{-2}, \qquad (4)$$

we find

$$\tfrac{1}{k_{\rm B}} S = d_{\rm eff} \cdot (\gamma_a - \gamma) \cdot \ln R + const., \qquad (5)$$

where the adiabatic polytropic index can be written in terms of the effective kinetic degrees of freedom, $\gamma_a = 1 + 2/d_{\rm eff}$ (Livadiotis 2015), so that, $\gamma_a \sim 5/3$, for $d_{\rm eff} \sim 3$, though these values can be affected by the magnetic field direction (Nicolaou & Livadiotis 2019) and magnitude (Livadiotis & Nicolaou 2021).

Then, the entropy gradient becomes

$$\frac{d}{d \ln R}\left(\tfrac{1}{k_{\rm B}} S\right) = d_{\rm eff} \cdot (\gamma_a - \gamma). \qquad (6)$$

Moreover, the gradient of the turbulent heating $E_t$ of the proton plasma (per mass), normalized by the thermal energy, is also equal to the deviation of the polytropic index $\gamma$ from its adiabatic value $\gamma_a$ (Verma et al. 1995; Vasquez et al. 2007; Livadiotis 2019), i.e.,

$$\frac{1}{k_{\rm B}T} \cdot \frac{dE_{\rm t}}{d \ln R} = \frac{d}{d \ln R}\left(\tfrac{1}{k_{\rm B}} S\right) = d_{\rm eff} \cdot (\gamma_a - \gamma). \qquad (7)$$

Combining the last two formulae together, we conclude in

$$\frac{1}{k_{\rm B}T} \cdot \frac{dE_{\rm t}}{d \ln R} = \frac{d}{d \ln R}\left(\tfrac{1}{k_{\rm B}} S\right) = d_{\rm eff} \cdot (\gamma_a - \gamma). \qquad (8)$$

The CMEs' sheath is highly turbulent compared to the ICMEs' ejecta or the solar wind plasma (Kilpua et al. 2019). This should be also captured in the thermodynamics of ICMEs. According to Eq.(1), the entropy is expected to increase and the polytropic index to decrease with increasing the turbulent heating. In particular, for $R\sim1$au, the entropic gradient is $\sim 3(\gamma_a - \gamma)$.

Ejecta is less turbulent CME regions; thus, thermodynamics is expected to recover at the original quasi-adiabatic process once the ejecta region is passed. The results shown in Figure 1d exactly verify this thermodynamic cycle: $\gamma \sim \gamma_a$ upstream of the I CME, $\gamma_a - \gamma \sim 0.26$ at the sheath, $\gamma_a - \gamma \sim 0.13$ at the ejecta, and again, $\gamma \sim \gamma_a$ at the end of the ICME. The corresponding entropic gradient $dS/dR$ is 0, +0.72 [$k_B$/AU], −0.36 [$k_B$/AU], and −0.36 [$k_B$/AU], respectively.

## 5. Discussion and Conclusions

Polytropic index, entropic gradient, and turbulent heating, all vary before, during, and after the passage of ICMEs at 1 au, revealing a cyclic behavior. In particular, the sheath is characterized by a low polytropic index, which deviates sharply from the adiabatic value prior to the ICME arrival at the spacecraft. Then, the index recovers and increases slightly during the ejecta passage and fully recovers to the pre-ICME levels after the ICME passes by.

Observational results of ICMEs at multiple spacecraft revealed that the sheath expands at a much faster rate (~factor of 2) than that of the ejecta within 1 au (e.g., Lugaz et al. 2020). The polytropic index characterizes compressions or expansions in the system and equivalently heat transfer processes. Although, dissipation of turbulence is not necessarily related to the polytropic process and is intrinsically different (Goldstein et al. 1995), turbulence levels are higher in the sheath, and the contribution of turbulence in altering the polytropic index cannot be ruled out. The observed faster expansion of the sheath may thus naturally correspond to an increased energy absorption, and thus resulting in a low polytropic index.